\documentclass[aps,amsfonts,amsmath,amssymb,prl,nofootinbib,reprint]{revtex4-2}
\usepackage{graphicx}
\usepackage{dcolumn}
\usepackage{bm}
\usepackage{tabularx} 
\usepackage{preamble}
\usepackage{enumerate}
\usepackage[font=small,labelfont=bf]{caption}
\usepackage{url}
\usepackage{pgfplots}
\usetikzlibrary{calc}
\usepackage{extarrows}
\usepackage{booktabs,subcaption,amsfonts,dcolumn}
\newcolumntype{d}[1]{D..{#1}}
\definecolor{refkey}{rgb}{0.9451,0.2706,0.4941}
\definecolor{labelkey}{rgb}{0.9451,0.2706,0.4941}
\def\MWCommentColor{ForestGreen}

\def\z2{$\mathbb{Z}_2$}

\definecolor{darkgray}{rgb}{0.33, 0.33, 0.33}
\usepackage{braket}

\newcommand{\Tr}{\mathop{\rm Tr}\nolimits}
\usepackage{amsthm}
\makeatletter
\def\l@subsection#1#2{}
\def\l@subsubsection#1#2{}
\makeatother
\pgfplotsset{compat=1.17}

\begin{document}

\title{A counterexample to the CFT convexity conjecture}
\author{Adar Sharon}
\email{asharon@scgp.stonybrook.edu}
\affiliation{Simons Center for Geometry and Physics, SUNY, Stony Brook, NY 11794, U.S.A.}
\author{Masataka Watanabe}%
\email{masataka.watanabe@yukawa.kyoto-u.ac.jp}
\affiliation{{Center for Gravitational Physics and Quantum Information (CGPQI), Yukawa Institute for Theoretical Physics, Kyoto University, Sakyo-ku, Kyoto 606-8502, Japan}}

\pagenumbering{arabic}
\pagestyle{plain}

\begin{abstract}
    Motivated by the weak gravity conjecture, \href{https://journals.aps.org/prd/abstract/10.1103/PhysRevD.104.126005}{[Phys. Rev. D {\bf 104}, 126005]} conjectured that in any CFT, the minimal operator dimension at fixed charge is a convex function of the charge. 
    In this letter we construct a counterexample to this convexity conjecture, which is a clockwork-like model with some modifications to make it a weakly-coupled CFT.
    We also discuss further possible applications of this model and some modified versions of the conjecture which are not ruled out by the counterexample.
\end{abstract}

\maketitle
\newpage


\section{Introduction}

The swampland program \cite{Vafa:2005ui} (see \cite{Palti:2019pca,vanBeest:2021lhn} for reviews) aims to find universal constraints on low-energy quantum field theories which can be uplifted to UV-complete theories of gravity. The result has been a series of conjectures which have gone through numerous stringent tests in various models, and to a deep understanding of relations between these conjectures.

An intriguing direction of study is to assume that swampland conjectures are also obeyed in any asymptotically AdS background, in which case the AdS/CFT duality maps them to statements which apply to arbitrary CFTs. This is very attractive since CFTs are simpler objects than quantum theories of gravity, and as such are much better understood. As a result, the CFT versions of swampland conjectures should be much easier to prove (or disprove). However, such claims require a significant leap of faith, since many swampland conjectures have mostly been studied in the context of weakly-coupled and weakly-curved gravitational theories. Indeed, so far only a handful of these types of ideas appear in the literature
-- among them are the CFT distance conjecture \cite{Baume:2020dqd,Perlmutter:2020buo}, the no-global-symmetries conjecture \cite{Harlow:2018jwu,Harlow:2018tng} and the cobordism conjecture \cite{Ooguri:2020sua,Simidzija:2020ukv}.

One of the most famous swampland conjectures is the weak gravity conjecture (WGC) \cite{Arkani-Hamed:2006emk}, which states that a $D$-dimensional effective low-energy gravitational theory with a $U(1)$ gauge field should include a particle whose mass and charge obey
\begin{equation}
    \frac{m}{q}\leq \alpha_D g (M_p^D)^{\frac{D-2}{2}}\;,
\end{equation}
with $\alpha_D$ some constant, $g$ the gauge coupling and $M_p^D$ the $D$-dimensional Planck mass.
It was originally motivated by demanding that all black holes are allowed to decay, and has since been put through many stringent checks. 

Several possible CFT duals of the WGC have been studied in the literature. The first attempt \cite{Nakayama:2015hga} focused on $d=4$ dimensional CFTs and demanded that black holes decay in $D=d+1=5$ dimensional AdS, which leads to the existence of an operator whose dimension $\Delta$ and charge $q$ obey 
\begin{equation}
    \frac{\Delta^2}{q^2}\leq \frac{9}{40}\frac{c_T}{c_V}\;.
\end{equation}
Here, $c_T$ and $c_V$ are the coefficients of the two-point functions of the energy-momentum tensor and the conserved current respectively.
Interestingly, this inequality has a simple CFT interpretation, as it also determines whether twists of double-twist operators have a convex or concave behaviour at large spin \cite{Komargodski:2012ek,Fitzpatrick:2012yx}.\footnote{The calculation in \cite{Nakayama:2015hga} can be extended to any $d>2$ and this match still occurs.}
Indeed, both questions essentially correspond to asking whether gravity is the weakest force in AdS. However, there are simple examples in which this inequality is not obeyed (like a free complex scalar in $d=4$), and so it is not the universal result one would have hoped for. Some additional discussions on the WGC and holography appear in \cite{Montero:2016tif,Heidenreich:2016aqi,Nakayama:2020dle,Montero:2018fns}.

A more recent attempt at deriving a universal CFT dual for the WGC was discussed in \cite{Aharony:2021mpc}. The motivation came from the positive binding energy conjecture \cite{Palti:2017elp,Heidenreich:2019zkl}, which is an alternative formulation of the WGC. In the simplest version of the conjecture, the authors assumed a unitary theory in $d>2$ with a single $U(1)$ symmetry, and defined $\Delta_{\text{min}}(q)$ to be the dimension of the operator of charge $q$ with the smallest dimension. Then the \textit{abelian convex charge conjecture} states that $\Delta_{\text{min}}(q)$ is convex, in the sense that there exists a charge $q_0$ of order 1 such that for any $n,m\in\mathbb{N}$, 
\begin{equation}
    \Delta_{\text{min}}((m+n)q_0)\geq \Delta_{\text{min}}(mq_0)+\Delta_{\text{min}}(nq_0)\;.
\end{equation}
This conjecture has been discussed and checked in many explicit examples in a series of recent papers \cite{Aharony:2021mpc,Dupuis:2021flq,Moser:2021bes,Watanabe:2022htq,Cuomo:2022kio}. In particular, some possible generalizations and refinements have appeared \cite{Antipin:2021rsh,Palti:2022unw}. 

As discussed in \cite{Aharony:2021mpc}, at large enough charge the conjecture is immediately obeyed in theories with a proper thermodynamic limit,\footnote{Free scalar theories or theories with moduli spaces of vacua are examples of theories with a sick thermodynamic limit \cite{Hellerman:2017veg}.
The spectrum becomes marginally convex in this case.} because of 
the universal scaling of operator dimensions at large global charge \cite{Hellerman:2015nra,Monin:2016jmo},
\begin{equation}\label{eq:large_charge}
    \Delta_{\text{min}}(q)\propto q^{\frac{d}{d-1}}+\text{(lower order in $1/q$)}\;.
\end{equation}
The content of the conjecture is thus in the statement that $q_0$, the charge at which convexity begins, is of order $1$. In order to find a counterexample, it is enough to find a theory in which $q_0$ can be made arbitrarily large.

In this letter we construct such a counterexample to the conjecture, and to most of the refinements in dimension $d>2$ which currently appear in the literature. The counterexample involves a weakly-coupled CFT in 3d, which we describe in section \ref{sec:counterexample}. In short, it is a 3d supersymmetric clockwork-like model \cite{Kaplan:2015fuy,Choi:2015fiu} with some modifications to make it a weakly-coupled CFT. We also discuss some generalizations and implications in section \ref{sec:implications}, and mention some refinements of the conjecture which are not ruled out by this counterexample. 

We emphasize that the CFT convexity conjecture was only motivated by the WGC, and is not equivalent to it. As a result, there are no immediate implications for any version of the WGC from this counterexample. In particular the counterexample does not have a weakly-coupled Einstein gravity dual.

Naively it might seem as though this letter is a step in the wrong direction. On the contrary, we would like to emphasize the rigor and simplicity of the CFT versions of the swampland conjectures, which can be disproven (and hopefully proven) with relative ease, making them worthy of further study. Specifically, it is clear that many CFTs \textit{do} obey the convexity conjecture, and so conceivably there is a small modification of the conjecture which should still hold. It would be interesting to find this modified version and attempt to prove it (see \cite{Orlando:2023ljh} for work in this direction).

\section{The counterexample}\label{sec:counterexample}

\subsection{The model}

We start with a 3d $\mathcal{N}=2$ $U(N)$ Chern-Simons (CS) theory at level $k$ coupled to $M$ matter multiplets $\Phi_i$ in the adjoint of the gauge group. The literature on these theories is vast, but we will require only the results from \cite{Gaiotto:2007qi,Chang:2010sg}, which we now review. We take the large-$N$ 't Hooft limit with $k\gg N\gg 1$. Then there exists a weakly-coupled fixed point with no superpotential for the $\Phi_i$'s, whose anomalous dimension $\gamma_\Phi$ is of order $-N^2/k^2$, sign included. Crucially, $\gamma_\Phi$ is both small and negative.

We can now deform this fixed point by a general superpotential of the form
\begin{equation}\label{eq:quartic}
    W=\sum_{i,j,k,l}\alpha_{ijkl}\Tr\left(\Phi_i\Phi_j\Phi_k\Phi_l\right)\;.
\end{equation}
Every $\alpha_{ijkl}$ is weakly relevant and so we can perform perturbation theory in the $\alpha_{ijkl}$'s. The $\beta$ functions at leading order in $1/k$ only get contributions from the superpotential, and one finds
\begin{equation}
    \beta_{ijkl}=4\gamma_\Phi\alpha_{i j k l}+\frac{1}{16 \pi^2} B_{(\underline{i}}{ }^r \alpha_{r \underline{j k l})}{}+\text{(higher loop)}\;,
\end{equation}
where $(\underline{ijkl})$ stands for cyclic symmetrization and
\begin{equation}
    B_i^j=\frac{1}{2} N^2 \alpha_{i k l m} \bar{\alpha}^{j k l m}\;.
\end{equation}

We now focus on a specific superpotential:
\begin{equation}\label{eq:exponential_model}
    W=4\sum_{i=1}^{M-1} \alpha_i\Tr\left(\Phi_i^3\Phi_{i+1}\right)\;.
\end{equation}
For generic $\alpha_i$'s, the theory now has two continuous global symmetries. One is the $U(1)_R$ R-symmetry, under which $\Phi_i$ has charge $1/2$.
The other symmetry is more interesting -- We call it $U(1)_E$ (for ``exponential''), under which $\Phi_i$ has charge $q_i=(-3)^{i-1}$. Importantly, $|q_i|$ grows exponentially with the index $i$.

Solving the equations above, one finds a fixed point at 
\begin{equation}
    |\alpha_i|^2=-\frac{8 \pi ^2  \gamma_\Phi}{N^2}\frac{\left(1-(-3)^i\right) \left(1-(-3)^{M-i}\right)}{\left(1+(-3)^M\right)}>0\;,
\end{equation}
where $i = 1,\, \dots,\, M-1$.
It is crucial that the RHS is always positive and hence the fixed point is unitary, and that there are no additional continuous symmetries at the fixed point. In addition, all of the $\alpha_i$'s are of order $O(\sqrt{-\gamma_\Phi})=O(N/k)$ and so the fixed point is still weakly coupled as expected.

\subsection{Lack of convexity}

Let us study convexity for the $U(1)_E$ symmetry at the fixed point. First, in order to have a single $U(1)$ symmetry, we weakly gauge the $U(1)_R$ by projecting out operators charged under it.\footnote{Note that by $U(1)_R$ we mean the naive R-symmetry, and not the ``true'' R-symmetry appearing in the superconformal algebra, which is some combination of $U(1)_R$ and other $U(1)$ symmetries determined by $F$-maximization \cite{Jafferis:2010un}.} This is done by adding a CS term to the gauge field with a very large level. Weakly gauging this symmetry explicitly breaks SUSY, but in a controlled way -- we are only removing some components of SUSY multiplets, but not changing the CFT data of the remaining operators. This step is technical and is required only in order to evade any issues with generalizations of the conjecture to more than one U(1) symmetry; equivalently, one could just consider operators which are charged only under $U(1)_E$ and not under $U(1)_R$.

The remaining gauge-invariant operators charged under $U(1)_E$ are single- or multi-trace operators such that the total R-charge vanishes. However, since the $\Phi_i$'s are the only fields charged under the $U(1)_E$ symmetry, and since we are only interested in the lowest-dimension operator at each charge, we can ignore other fields (like $F_{\mu\nu}$). Denoting the components of $\Phi_i$ by $\phi_i,\psi_i$ which have R-charge $1/2,-1/2$ respectively, we are thus only interested in single- or multi-trace operators made of $\phi_i,\psi_i$ and their conjugates
such that the total R-charge vanishes. Since the model is weakly-coupled, we are assured that these will be the only operators which are of interest to us. For example, one can take 
\begin{equation}\Tr\left(\phi_{i_1}...\phi_{i_n}\bar\phi_{j_1}...\bar\phi_{j_m}\psi_{k_1}...\psi_{k_p}\bar\psi_{l_1}...\bar\psi_{l_r}\right)\;,
\end{equation}
with $n-m-p+r=0$. We must also rescale the $U(1)_E$ charge such that the minimal charge of all gauge-invariants is $1$, so that now $\Phi_i$ has charge $q_i=\frac{(-3)^{i-1}}{2}$.

Now we show that there is no convexity for the $U(1)_E$ symmetry for all $q_0$ which obey $q_0<\frac{3^{M-3}}{2}$ and such that the lowest-dimension operators at charges $q\leq 9q_0$ consist of less than $N$ fields.\footnote{In particular, by taking $N,M$ to be large, all $q_0$ up to some arbitrarily large value can be made to obey these assumptions.} 
Assuming such a $q_0$, there exists some gauge-invariant operator which has minimal dimension at charge $q_0$ of the form discussed above. 
Since the fixed point is weakly-coupled, its dimension is\footnote{Mixing does not affect this result since the number of operators which can mix does not depend on $N,k$.} 
\begin{equation}
    \Delta_{\text{min}}(q_0)=\frac{
    \#\text{scalars}}{2}+\#\text{fermions}+O(N/k)\;.
\end{equation}
Next consider operators of charge $q=9q_0$. We can construct an operator at this charge by taking the lowest dimension operator at charge $q_0$ and shifting all of the indices of the fields by $2$ (in the example above this corresponds to $i_s\mapsto i_s+2$ for all $s$, etc.). Since we assumed $q_0<\frac{3^{M-3}}{2}$, the lowest-dimension operator with charge $q_0$ consists only of fields whose indices are at most $M-2$, and so shifting these indices by 2 is allowed. 
Since this new operator is constructed using the same number of fields as the operator for $q_0$, it has the same dimension up to corrections of order $N/k$. We thus find
\begin{equation}
    \Delta_{\text{min}}(9q_0)\leq\Delta_{\text{min}}(q_0)+O(N/k)\;,
\end{equation}
while convexity requires
\begin{equation}
    \Delta_{\text{min}}(9q_0)\geq 9\Delta_{\text{min}}(q_0).
\end{equation}
Combined with the unitarity bound $\Delta_{\text{min}}\geq \frac{1}{2}$, we find that the spectrum is not convex for large enough $k$. 
We can now make $k,N,M$ arbitrarily large (while keeping $k\gg N\gg M$), 
thereby parametrically delaying convexity as much as we would like and providing a counterexample to the convexity conjecture.

\begin{figure}[t]
\includegraphics[width=0.4\textwidth]{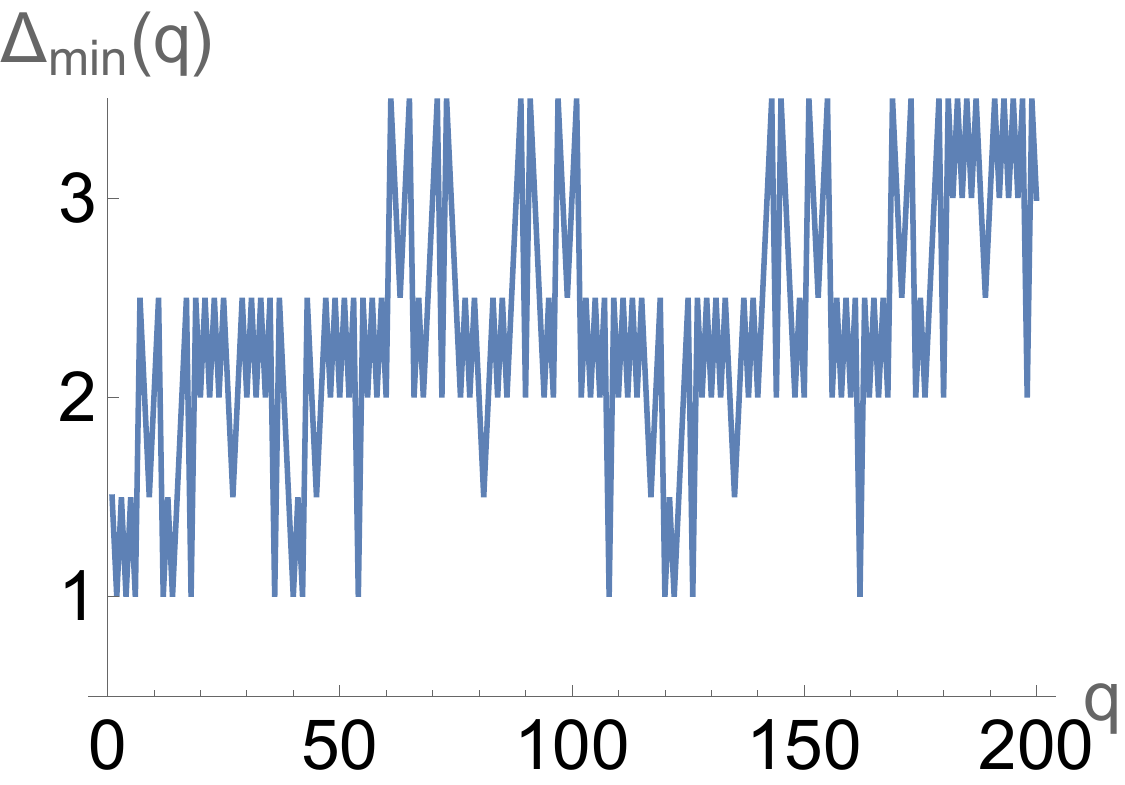}
\caption{$\Delta_{\text{min}}(q)$ for small charges in the counterexample.}
\label{fig:fig}
\centering
\end{figure}

For completeness we present $\Delta_{\text{min}}(q)$ for $q\leq 200$ and large enough $k,N,M$ in figure \ref{fig:fig} to make the lack of convexity apparent.

\section{discussion}\label{sec:implications}

Some generalizations of the convexity conjecture have appeared in the literature, and we start by remarking on their validity. 
First, \cite{Aharony:2021mpc} mentions a version of the conjecture which applies only to scalar operators, but this conjecture also fails due to the same counterexample discussed above. 
Some generalizations have also appeared for theories with more than one $U(1)$ symmetry and for nonabelian symmetries \cite{Aharony:2021mpc,Palti:2022unw,Antipin:2021rsh}, but since the case with a single $U(1)$ symmetry is a special case of these the counterexample rules them out as well.

Finally, the 2d version of the conjecture was discussed in \cite{Palti:2022unw}. This version is different from the higher-dimensional version, since there it was already shown that convexity can be parametrically delayed. However, \cite{Palti:2022unw} provided an upper bound on the delay in convexity which is related to the level of the corresponding WZW theory. This version of the conjecture was proven explicitly and so it still stands. 

An obvious question is whether there is a small refinement of the conjecture which rules out the counterexample discussed above. 
One option appears in \cite{Aharony:2021mpc}, where $q_0$ is determined by the charge of the lowest-dimension charged operator. 
Computing the anomalous dimensions of various single-trace operators, we find that this sets $q_0\sim 3^{M-1}/2$ at large $M$ and so this version of the convexity conjecture is not ruled out by the counterexample (however, $q_0$ can still be made to be arbitrarily large).\footnote{We thank Ofer Aharony for discussions on this point.} 
Another option is to allow convexity to be delayed up to some upper bound which depends exponentially on the number of fields in the theory. 
Explicitly, one can conjecture that $q_0\leq e^{N_D}$, where $N_D$ is some measure of the number of degrees of freedom in the theory (e.g.~$c_T$, the $c$ anomaly in 2d, the $a$ anomaly in 4d, or the $F$-coefficient in 3d). 
Another option would be to apply the conjecture only to theories with a weakly-coupled Einstein gravity dual. We will not speculate on such refinements further.

We note that it seems that $\mathcal{N}=2$ SUSY is important in order for the counterexample to remain tractable (even though we have effectively removed SUSY by weakly gauging the $U(1)_R$ symmetry). One could in principle consider a $3d$ $\mathcal{N}=1$ or non-SUSY version of this counterexample (see \cite{Gomis:2017ixy,Bashmakov:2018wts} for recent discussions of such CS theories with adjoints). 
However, a practical issue is that in addition to a potential which mimics the behaviour of the superpotential \eqref{eq:exponential_model}, there are many other terms that can appear under the RG flow. For example, one can study the $3d$ $\mathcal{N}=1$ CS-matter theory with superpotential \eqref{eq:exponential_model}, but now the RG flow also generates interactions of the form
\begin{equation}
    \alpha_{ij}\Tr\left(|\Phi_i|^2|\Phi_j|^2\right)\;,
\end{equation}
and one must study whether there exist fixed points for all of these couplings as well, which complicates the analysis considerably.

Next we discuss possible implications to the WGC. As noted in \cite{Aharony:2021mpc}, the charge convexity conjecture in no way implies the WGC (or the other way around), and so this counterexample cannot be explicitly used to learn much about quantum gravity. At best it can be used to perhaps motivate an analogous gravitational construction, but it is not clear how to do this. 
First, the CFT we constructed does not have a weakly-coupled Einstein gravity  dual. 
Second, the main mechanism used here is the same as in the clockwork model, which is not a new idea in the context of swampland. However, it is possible that the counterexample suggests a closer look at clockwork-like models in the context of the WGC as well.

We end by commenting on some other possible applications of the model we discussed.
First, the model may be interesting to study in the context of the large-charge expansion, which has been shown to be approximately valid down to surprisingly low charges in simple examples even when taking into account only the first few leading terms in the expansion \cite{Banerjee:2017fcx, Hellerman:2017efx, Hellerman:2018sjf, Banerjee:2019jpw}. 
However, in our model this will almost definitely not occur, since $\Delta_{\text{min}}(q)$ is an erratic function of $q$ for small enough charges (see figure \ref{fig:fig}).
It would thus be interesting to better understand when the large-charge expansion can be safely extrapolated down to small charges. 
It is possible that one can bound the validity of the expansion at small charge by using some function of the number of degrees of freedom in the CFT, as discussed above.
The class of theories discussed above with more general superpotentials \eqref{eq:quartic} may also have many possible applications as a weakly-coupled conformal manifold. 
In particular, one could try to define an SYK-like theory in 3d by averaging over this conformal manifold (see \cite{Chang:2021fmd} for another attempt in 3d). 
Then observables like the chaos exponent can be extracted using the formalism discussed in \cite{Berkooz:2022dfr,Kalloor:2023hnq,Berkooz:2021ehv}, which generalizes SYK-like models to those with general CFTs deformed by disorder.

\section*{Acknowledgements}
We would first like to thank Ofer Aharony and Eran Palti for many interesting discussions, and especially for their patience and for helping to disprove many previous attempts at counterexamples. 
We are also grateful to Simeon Hellerman for helpful discussions which initialised the project.
We would also like to thank Shota Komatsu, Miguel Montero, Domenico Orlando and Sridip Pal for illuminating discussions. 
MW is grateful to the conference ``Large charge aux Diablerets'' where this work began, and AS is grateful to the Abu Dhabi meeting on theoretical physics where this work was finished. 
MW is supported by Grant-in-Aid for JSPS Fellows (No. 22J00752).

\bibliographystyle{JHEP}
\bibliography{ref,main}

\end{document}